\documentclass[aps,pre,twocolumn,showpacs,preprintnumbers]{revtex4}
\usepackage{times,mathptmx}
\usepackage{epsfig}
\usepackage{amssymb}

\begin{document}

\title{Analysis of the SGA  method for obtaining energy  spectra}
\author{
Patricio Cordero and Jamil Daboul\footnote{On Sabatical 
leave from the Physics Department, Ben Gurion University of the Negev,
84105 Beer Sheva, Israel (e-mail: daboul@bgu.ac.il}
}
\affiliation{Departamento de F\'{\i}sica, Universidad de Chile,
Santiago, Chile}
\pacs {03.65.Fd, 03.65.-w,  02.20.-a}

\begin{abstract}

We analyze and clarify how the SGA (spectrum generating algebra) method has
been applied to different potentials.  We emphasize that each energy level
$E_\nu$ obtained originally by Morse belongs to a {\em different}
${\mathfrak {so}}(2,1)$ multiplet. The corresponding wavefunctions $\Psi_\nu$
are eigenfuntions of the compact generators $J^\nu_0$ with the same
eigenvalue $k_0$, but with different eigenvalues $q_\nu$ of the Casimir
operators $Q$. We derive a general expression for all effective potentials
which have $\Psi_{\lambda_\nu,\nu+m}(r) \propto 
(J_+^\nu)^m ~\Psi_{\lambda_\nu,\nu}(r)$
as eigenfunctions, without using super-symmetry formalism. The different
actions of SGA is further illustrated  by two diagrams.

\end{abstract}

\maketitle

\section{Introduction}

Pauli \cite{pauli} in 1926 was the first person who calculated the energy
spectrum of a Hamiltonian {\em algebraically}. He did it for the hydrogen
atom. Since then his procedure was followed by many people: It is
essentially based on relating the {\em total} Hamiltonian $H$ to the Casimir
operator of the {\em symmetry or degeneracy algebra}, whose generators
commute with $H$. (The symmetry algebra of the N-dimensional hydrogen atom
is usually identified differently for different energies $E$. By replacing
the Hamiltonian $H$ by its eigenvalues, one obtains ${\mathfrak
{so}}(N+1)$,~ ${\mathfrak {e}}(N)$ and \, ${\mathfrak {so}}(N,1)$, ~for $E<0,~ E=0$
and $E>0$, respectively \cite{schiff,sud,nieto}. However, by keeping $H$ as
operator, one obtains infinite-dimensional Kac-Moody loop algebras, of standard
type for even $N$ and of twisted type for odd $N$ \cite{dsd,dd}).

The hydrogen atom and the isotropic oscillator have infinite number of
states for each eigenvalue of angular momentum $\ell$. Their symmetry algebras for the bound
states are compact and therefore the raising and lowering operators of these algebras can only generate a {\em finite} number of states with different values of $\ell$, for each degenerate energy eigenvalue 
$E_n$. To relate the infinite number of states for a fixed $\ell$ it is necessary to use non-compact
algebras. The smallest appropriate noncompact algebra is ${\mathfrak {so}}(2,1)$, because it has an infinte-dimensional representations which are 
bounded from below, and are denoted by $D^+(\lambda)$.
The  ${\mathfrak{so}}(2,1)\,$ is generated by three generators $K_i$ which commute as follows
\begin{eqnarray}
~ [K_0, K_1] &=& i  K_2~, \cr
~  [K_2, K_0] &=&  i  K_1\,,  \label{Kcr}  \\
~  [K_1, K_2] &=& -i  K_0\,.  \nonumber
\end{eqnarray}
The $K_0$ is the compact generator, while $K_1$ and $K_2$ are the noncompact
ones. Once a simultaneous eigenstate of $K_0$ and $Q$ is found, then one can generate an
infinite number of states by applying the raising and lowering operators $K_\pm$ repeatedly.
Because of this property ${\mathfrak {so}}(2,1)$ has been called a {\em spectrum-generating algebra} (SGA).

The idea of using the SGA was very
popular in the 1960's and 1970's and numerous papers \cite{many} investigated different
approaches and potentials. In fact, Wybourne devoted a whole chapter of his popular book
\cite [Chap. 18]{wy} to describe how one obtains the energy levels of
various systems from the ${\mathfrak {so}}(2,1)$ SGA. 

In particular, ${\mathfrak {so}}(2,1)$ has been applied to obtain 
the S-wave bound states for a important subclass of Natanzon 
potentials  \cite{cs}, which has 
only a finite number of bound states, such as the Morse potential. 
The question arises, {\em where are the other infinitely many energy levels that
can be generated  by $K_\pm$ }?    

The purpose of the present paper is to show that in the many applications of
${\mathfrak {so}}(2,1)$ involving Natanzon-type potentials~\cite{natanzon}, the raising and lowering
operators $K_\pm$ were actually never used to generate the finite number of
bound states for $\ell=0$. Instead, it turned out that for every energy
level, a different  ${\mathfrak{so}}(2,1)\,$ representation was used .

For clarity, we shall concentrate on the Morse potential,
\begin{equation}
V_M(r) := V_0 \left[e^{-2(r-r_0)/a}- 2 e^{-(r-r_0)/a} \right ]~.
\label{V}\end{equation}
which is the simplest example  of a Natanzon potential which has a finite
number of bound states for each $\ell$.  The potential (\ref{V}) was introduced
by Morse in 1929  to obtain the vibrational levels of diatomic
molecules \cite{morse}.  He obtained only the S-wave solutions, by explicitely solving
the Schr\"odinger equation.   The Morse potential has
since become popular, especially among chemists, because it allows
disintegration of diatomic molecules, in contrast to the shifted
harmonic-oscillator potential, $ (k/2) (r-r_0)^2$.

In 1970 Cordero and Hojman \cite{ch} (hereafter will be quoted as CH)
reproduced algebraically the finite S-wave energy spectrum for the {\em Morse
potential} \cite{morse}, by using  ${\mathfrak{so}}(2,1)\,$. 
Therefore we shall analyze this paper in particular, and show that CH did not use a single
${\mathfrak{so}}(2,1)\,$ representation to get the enegy levels, 
but actually a different  ${\mathfrak{so}}(2,1)\,$ representation 
for each energy level. 
We  explain how CH succeeded nevertheless in obtaining the correct energy
spectrum. This should be worthwhile, since the CH paper succeeded for the
first time to produce a finite number of bound states from a SGA.

In Sec. II we review the SGA method for the 3-D oscillator and in Sec. III we review the
paper of CH on the Morse potential, and point out the basic difference in the two cases. 
We shall see that $K_0$ commutes with $H_\ell$ in the case of the oscillator, but it does not commute with $H_S$ in the Morse case;
in the latter case the Casimir operator $Q$ commutes with $H_S$.
In Sec. IV we derive the effective potentials $V_M(k_m,r_m;r)$ (see (\ref{Vm}) below),  which have
the functions 
$\Psi_{\lambda_\nu,\nu+m}(r) \propto (J_+^\nu)^m ~\Psi_{\lambda_\nu,\nu}(r)$ as eigenstates, where
$\Psi_{\lambda_\nu,\nu}(r)$ are  the wavefunctions of the Morse potential (\ref{V}) and $J_+^\nu$
are raising operators. Finally, in Sec. V we give a summary. 

\section{Energy spectrum of the 3-D harmonic oscillator}

In this section we review the derivation of the energy spectra for
partial-wave Hamiltonians $H_\ell$ of the 'generalized' (by adding the $\epsilon$)
harmonic oscillator
\begin{eqnarray}
H_\ell &=& - \frac {\hbar^2}{2M} \frac{d^2}{dr^2} + \frac {(\ell(\ell+1)+\epsilon) \hbar^2}{2Mr^2} + \frac k 2 r^2~,\cr && \cr
& =& 
\frac {\hbar^2}{2Ma^2} \left(- a^2\frac{d^2}{dr^2} + \frac {\alpha_\ell ~a^2} {r^2} 
+ \frac {a^4 Mk}{\hbar^2} \frac{r^2}{a^2}  \right) \cr
&& \cr  
& =& 
{\cal E} \left(- \frac{d^2}{dy^2} + \frac {\alpha_\ell} {y^2} + \left(\frac {a^2 M \omega }{\hbar}\right)^2 y^2  \right) 
\label{hl}\end{eqnarray}
for all $-\infty < k <  \infty$, where we use the notation
\begin{eqnarray}
\omega &\equiv& \sqrt{|k|/M} ,\quad  y\equiv  r/a,\cr
\alpha_\ell &\equiv& \ell(\ell+1)+\epsilon, \quad {\quad \mbox{ and}\quad} \cr 
{\cal E} &\equiv&  \frac {\hbar^2}{2Ma^2} 
\label{ec}\end{eqnarray}
Note that $\alpha_\ell$ are dimensionless constants. We introduced  the scaling
factor $a$ which has the dimension of length, and whose value will be
determined below in (\ref{s}). Thus, $y$ becomes a dimensionless variable,
and ${\cal E}$ has the dimension of energy.

Note that we defined $H_\ell$ for the attractive oscillator $k>0$ and also for repulsive
oscillator $k<0$\,. Usually the oscillator is only studied for
attractive case, but $k$ was defined in \cite{dab} for all real values
of $k$, in order to study the contraction of the algebra ${\mathfrak
{su}(2)}$ to the Euclidean algebra ${\mathfrak {e}}(2)$ or to the
Heisenberg algabra ${\mathfrak h}(3)$, as $k\to 0$.

\subsection{Realization of  ${\mathfrak{so}}(2,1)\,$ ~generators for the oscillator}

It is easy to check that the following three generators satisfy the 
commutation relations (\ref{Kcr}) of  ${\mathfrak{so}}(2,1)\,$ 
\begin{eqnarray}
K_0 (\alpha) &:= & -\frac{d^2}{dy^2} + \frac {\alpha}{y^2} + \frac {y^2}{16}  \cr 
K_1 (\alpha) &:= & -\frac{d^2}{dy^2} + \frac {\alpha}{y^2} - \frac {y^2}{16} = K_0 -\frac {y^2}{8}  \label{kwy} \\ 
K_2 &:=& =\frac {-i} 2 \left(y \frac{d}{dy} + \frac{1}{2} \right)~,   \nonumber
\end{eqnarray}
and thus yield different realizations of  ${\mathfrak{so}}(2,1)\,$ for every value of the constant $\alpha$. 
(The generators in
(\ref{kwy}) follow from those in \cite[Eq.(18.7)]{wy} by multiplying
$K_0$ and $K_1$ by a minus sign, which leaves the commutation relations unchanged, and then  
by replacing $\alpha$ by $-\alpha$.
The above generators are so constructed, that if they are applied to an eigenfunction of $K_0$,
then their Casimir operator is related to $\alpha$, as follows 
\begin{equation}
Q :=  K_0^2 - K_1^2- K_2^2 =  \left(\frac {\alpha}{4} - \frac {3}{16} \right) ~I  =:
q (\alpha)\,  I\,,
\label{Qw}
\end{equation}
where $I$ is identity operator.
If we factorize $q$ as follows
\begin{equation}
q(\alpha) := \lambda(\lambda-1) =  \frac {\alpha}{4} - \frac {3}{16}  \,,
\label{Ql}\end{equation}
and solve the quadratic equation in (\ref{Ql}), we obtain for $\lambda$ the values
\begin{eqnarray}
\lambda_\pm (\alpha) &=& \frac{1}{2} \pm \frac{1}{2} \sqrt{1+4q} \nonumber \cr
&=& \frac{1}{2} \pm \frac{1}{2} \sqrt{\alpha+ 1/4}, {\quad \mbox{so that}~~} \alpha \ge -1/4\,.
\label{la}
\end{eqnarray}
It is interesting to note that the realization (\ref{kwy}) for $\alpha=0$ also
yields the SGA of the one-dimensional harmonic oscillator. In this case,
(\ref{la}) yields the well-known values $\lambda_-=1/4$ and $\lambda_+=3/4$, which define
the two infinite-dimensional representations, the Fock states $|{2m}\rangle$
and $|{2m+1}\rangle,~m=0,1,2,\cdots$, respectively \cite{vw}.

\subsection{Relating $H_\ell$ to the generators of  ${\mathfrak{so}}(2,1)\,$}

The realization (\ref{kwy}) is suitable for generating the
oscillator states. In fact, $H_\ell$ becomes proportional
to $K_0$ for $k>0$ or to $K_1$ for $k<0$, if we scale the coefficient of
$y^2$ in (\ref{hl}) to $1/16$, {\em i.e.}
\begin{equation}
\sigma \equiv \frac {a^4 M^2 \omega^2 }{\hbar^2} \Rightarrow \frac 1 {16}
\end{equation}
which is equivalent to choosing $a$, as follows
\begin{equation}
a = \left(\frac {\hbar^2}{16\, M|k|}\right)^{1/4} = \left(\frac {\hbar}{4M\omega}\right)^{1/2}~.
\label{s}
\end{equation}
With this choice of $a$, we obtain
\begin{equation}
{\cal E}= \hbar^2/(2Ma^2)= 2 \hbar \omega ~.
\label{ecw}\end{equation}
Hence (\ref{hl}) becomes 
\begin{eqnarray}
H_\ell &=& {\cal E} \left(- \frac{d^2}{dy^2} + \frac {\alpha}{y^2} + {\rm sgn} (k) 
 \frac {y^2}{16}\right)~,  \cr
&& \cr
&=& \left\{\begin{array}{ll}  2 \hbar \omega ~K_0  &{\qquad \mbox{ for}\quad} k>0 {\quad \mbox{ and}\quad} \cr 
& \cr
2 \hbar \omega~ K_1 &{\qquad \mbox{ for}\quad} k< 0 \end{array}   \right.
\label{hk}\end{eqnarray}

Hence, the eigenfunctions of $K_0$ will be the eigenfunctions of
$H_\ell$ for $k>0$. Since the eigenvalues of $K_0$ are given by
\cite{wy,bf} $\nu+\lambda , ~\nu=0,1,2,\cdots$, it follows
that the energy spectrum for the $\ell$-partial wave
is given, for $\alpha_\ell = \ell(\ell+1)+\epsilon$, by \cite[\S 36]{lal} and \cite{wy}
\begin{eqnarray}
E_{\nu,\ell} &=& 2 \hbar \omega (\nu+ \lambda_+ (\alpha_\ell)) ~~\qquad \nu =0,1,2,\cdots~,   \cr
&& \cr
&=& \hbar \omega (2\nu+ 1 + \sqrt{(\ell+1/2)^2+\epsilon})   
\end{eqnarray}
which tends, in the limit $\epsilon \to 0$ to
\begin{equation}
 E_{\nu,\ell} = \hbar \omega (n + 3/2), {\quad \mbox{where}\quad} n:=2\nu+\ell
\label{eosc}
\end{equation}
It can be seen that the shift in $\lambda_\ell$ in the eigenvalues of
$K_0(\alpha_\ell)$, due to $\epsilon$, is also multiplied by the factor
$2\hbar \omega$.  Note that for a fixed $\ell$ the energy levels
$E_{\nu,\ell}$ increase by $2\hbar \omega$ rather than $\hbar
\omega$.  

\begin{figure}[htb]
\epsfig{file=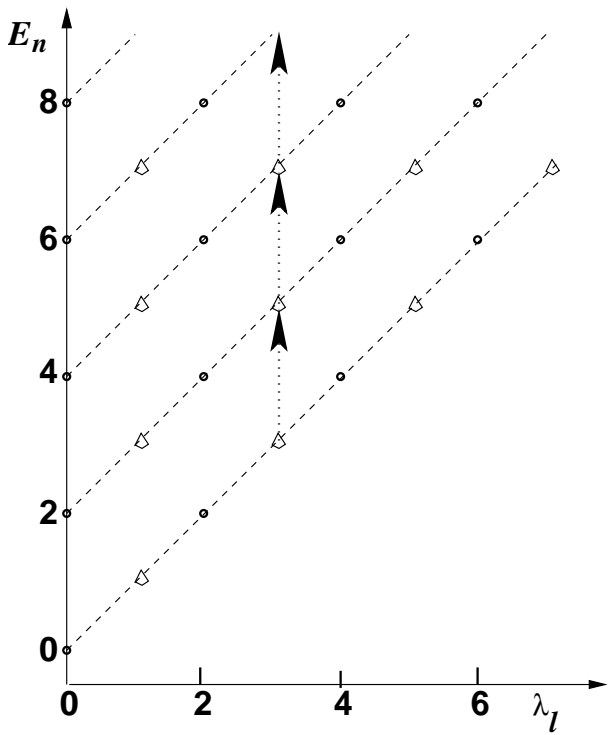,width=8.0cm,angle=0}
\caption{ The eigenstates of the harmonic oscillator for a given $\ell$
belong to a single irreducible representation of ${\mathfrak {so}}(2,1)$
which is
 characterized by $\lambda_\ell$\,. The excited states
for a given $\ell$ can be obtained by applying the raising operator
$K_+(\alpha_\ell)$, as illustrated by the vertical arrows.
\label{fig:Fwy} 
}
\end{figure}

In the case of the harmonic oscillator  
all the energy levels for a given $\ell$ belong to a single 
representation of ${\mathfrak{so}}(2,1)\,$,
so that all eigenfunctions $\Psi_{\ell,\nu}$ can be 
obtained by applying powers of the raising operators $K_+(\alpha_\ell)$ on 
the ground state $\Psi_{\ell,0}$\,, as illustrated in Fig.~\ref{fig:Fwy}.

\section{Algebraic derivation of the Morse spectrum}

In CH the authors reproduced the S-wave energy spectrum algebraically by
using ${\mathfrak{so}}(2,1)\,$. We now show that, in contrast to the
oscillator, one cannot obtain the S-wave spectrum of the Morse potential by
using one representation of ${\mathfrak{so}}(2,1)\,$. In fact, it turns out,
that for each energy level $E_\nu$ one needs a different realization
$J_i^\nu$ (see (\ref{knu}) below) of the ${\mathfrak{so}}(2,1)\,$
generators. This fact is essentially implied by the formulas they used, but
it was never stated clearly neither in the above paper, nor in subsequent
papers on the Natanzon potentials \cite{cs}. Before explaining their
procedure, we introduce new notation and also note some changes of notation
from that in CH.
 
\subsection{New notation}

We try to make the present paper self-contained. However, if the reader
likes to consult the original paper of CH, he should note the following
changes of notations and definitions which we have made, so that it becomes
easier to check the dimensions in the formulas and to simplify some of them.

In CH the Morse potential was written as $V_{old} (r):= D
\left[e^{-2a_{old}(r-r_0)}- 2b e^{-a_{old}(r-r_0)} \right ] $.
In this paper we define $a:= 1/a_{old}$, so that $a$ has the dimension 
of {\em length}. We also
replace the unnecessary parameter $b$ by 1, by defining $V_0 = D b^2$ and adjusting 
the value
of $r_0$. With the new definition the potential $V(r)$ has its minimum at $r=r_0$. 

We find it quite useful to introduce the following dimensionless constant
\begin{equation}
k_0 \equiv \frac 1 \hbar  \sqrt{2Ma^2V_0} = \sqrt{V_0/{\cal E}}~, {\quad \mbox{so that}~~} V_0={\cal E} k_0^2~,
\label{k0}\end{equation}
where ${\cal E}$ is defined by the expression (\ref{ec}).
We shall see that $k_0$ is equal to the eigenvalue of $K_0$, when it acts on the $S-$ wave solutions.

The following dimensionless exponential function  is also very useful  
\begin{equation} 
h (r):= k_0 e^{-(r-r_0)/a}
\label{h}\end{equation}
Note that $h (r_0):= k_0 $ and that our $h(r)$ in (\ref{h}) 
is equal to {\em twice} the $h_{old} (r)$ in CH; this redefinition 
simplifies many formulas, by making factors of $1/2$ and $1/4$ unnecessary.

We can now write the Morse potential in terms of $h(r)$ simply, as follows
\begin{equation}
V_M(r) = V_0  \frac {h(r)}{k_0} \left(\frac {h(r)}{k_0}- 2\right)
= {\cal E}~ \left(h(r)^2 -2k_0 h(r) \right)
\label{ve}\end{equation}

\subsection{Relating the partial Hamiltonian $H_S$ to $J_0$}

The following three operators were defined in CH,
which depend on a parameter $\beta$, which the authors  called 
$-E/{\cal E}$,
\begin{eqnarray}
 J_0 (\beta) &:= & \frac 1 {2 h(r)} \left[-  a^2 \Delta_r + \beta +  h^2(r) \right] 
 \cr
 J_1 (\beta) &:= & J_0 -  h  \label{j012} \\ 
 J_2 &:=& i \left(a\frac{d}{dr} + \frac a r -\frac{1}{2} \right)~,  \nonumber
\end{eqnarray}
where $h(r)$ is the exponential function (\ref{h}) and
$\Delta_r $ is the radial part of the Laplacian $\Delta$, {\em i.e.}
\begin{equation}
\Delta_r := \frac{d^2}{dr^2}+ \frac 2 {r} \frac{d}{dr} = \left(\frac{d}{dr}+ \frac 1 r \right)^2  
\label{dr}\end{equation}
These $J_i$ commute exactly as the $K_i$ in (\ref{Kcr})
and hence they also yield a realization of the non-compact  ${\mathfrak {so}}(2,1)$ algebra. 

The realization $J_i$ in (\ref{j012}) is defined  such that it acts on the radial part  
$ \Psi_{\nu,\ell} (r)=R_{\nu,\ell}/r$  of the wave function 
$\Psi_{\nu,\ell} ({\bf r})=R_{\nu,\ell}/r Y_{lm} (\theta,\varphi )$.
It can be transformed into the realization  $K_i$ of  ${\mathfrak{so}}(2,1)\,$ which acts on $R(r)$, as follows
\begin{equation}
K_i := r ~J_i ~\frac 1 r
\label{trans}\end{equation}
The Casimir operator $Q$ of the $J_i$ is related to the parameter $\beta$ by
\begin{equation}
Q:= J_0^2 - J_1^2-J_2^2 = (\beta -\frac{1}{4}) I=: q(\beta)~ I ~, 
\label{Q}
\end{equation}
so that we may replace $\beta$ in (\ref{j012}) by $q+1/4$. 
Note that the above expression for 
$q(\beta)$ for the  $J_i$ in
(\ref{j012}) is different from that for $q(\alpha)$ in (\ref{Qw}).  The corresponding 
\begin{equation}
\lambda_\pm (\beta) = \frac{1}{2} \pm \frac{1}{2} \sqrt{1+4q} 
=\frac{1}{2} \pm \sqrt{\beta}, 
\label{la2}
\end{equation}
require that $\beta \ge 0$ in order that $\lambda_\pm $ to be real.

The {\em `S-wave Hamiltonian'} for the Morse potential
\begin{equation} 
H_S :=  - \frac {\hbar^2} {2M} \Delta_r  + V_M(r) 
\label{hs}\end{equation}
is related to $J_0$, as follows 
\begin{equation}
H_S =   {\cal E} \,[\,2\,h(r)\, (J_0(\beta)-k_0)-\beta] ~. \label{hsj}
\end{equation}
Note that $J_0$ does not commute with $H_S$, because $J_0$ does not commute with $h(r)$.

\subsection{Condition on the wavefunctions $\Psi_\nu (r)$}

Let  $\Psi_\nu (r)$,  denote 
the eigenfunction of $H_S$ associated to the  eigenvalue
 $E_\nu$, for all $\nu$\,, $\nu=0,1,2,\dots,\nu_{max}$\,,  {\em i.e.}
\begin{equation}
[H_S-E_\nu]~\Psi_\nu (r) = [{\cal E} (2h(J_0-k_0)-\beta)-E_\nu]~\Psi_\nu (r) =0~. \label{sch}
\end{equation} 
This equation can be satisfied, {\em iff} $\Psi_\nu (r)$ are eigenfunctions of $J_0$,
with the {\em same} (!) eigenvalue $k_0$ for all the allowed $\nu$, {\em i.e.}
\begin{equation}
J_0~\Psi_\nu (r) = k_0~\Psi_\nu (r) ~, {\qquad \mbox{ for}\quad}  \nu=0,1,2,\cdots,\nu_{max}.
\label{c1}\end{equation}
{\em and} if  for every $E_\nu$ we choose $\beta_\nu$, such that
\begin{equation}
-\frac {E_\nu} {\cal E} =\beta_\nu = q_\nu +\frac{1}{4} =\lambda_\nu (\lambda_\nu -1)+\frac{1}{4}=(\lambda_\nu -1/2)^2~.
\label{c2}\end{equation}
But since $k_0$ is an eigenvalue of the compact generator, it must be related to 
$\lambda_\nu$ as follows \cite{bf,wy}
\begin{equation}
k_0 = m_\nu + \lambda_\nu~,
\label{condm}\end{equation}
where $m_\nu$ is some integer.

To fix the integers $m_\nu$, we proceed as follows: First, we order for definitness the
$E_\nu$, such that $E_0 < E_1< \cdots < E_{\nu_{max}}$~.
By noting (\ref{c2}) we conclude that the ground state $\Psi_0$ must belong to the highest possible
value of $\lambda_\nu$ consistent with the condition (\ref{condm}). Hence, we must choose $m_0=0$ for
$\lambda_0$. Following similar arguments, we finally obtain
\begin{equation}
k_0 = \nu + \lambda_\nu~, {\quad \mbox{or} \quad} \lambda_\nu = k_0 - \nu ~.
\label{condf}\end{equation}
To obtain (\ref{condf}) we implicitly assume that every permissible solution 
 is a physical eigenfunction. 
Substituting this expression for $\lambda_\nu$ into (\ref{c2}), we obtain the same energy spectrum 
for the S-wave bound states as  in \cite{ch}
\begin{eqnarray} 
E_\nu (\ell=0) &=& - {\cal E} \left[k_0 -\frac{1}{2}-\nu
 \right]^2 \cr && \cr
&=& - \frac  {\hbar^2} {2Ma^2} \left[\frac{\sqrt {2Ma^2V_0}}{\hbar} -\frac{1}{2} -\nu \right]^2 ,
\label{enu}
\end{eqnarray}
where
\begin{equation}
\nu=0,1,2\cdots,\nu_{max}=\lfloor k_0-1/2\rfloor_- \,. 
\end{equation}
This spectrum was first obtained by Morse \cite{morse} by solving the
Schr\"odinger equation. The value of $\nu_{max}=\lfloor k_0-1/2\rfloor_- $,
where we use the notation $\lfloor x\rfloor_-$ to denote 
the largest integer which is {\em smaller} (not equal] to
of $x$, because $s:= k_0 -\frac{1}{2}-\nu > 0$ in order for the solution to
be normalizable \cite{morse}.  (see also the comment in Sec.\ref{trad} ). 
Hence, for $k_0 \le 1/2$ there are no bound states.

It is important to note that the energy levels $E_\nu$ in (\ref{enu}) do not
depend on $r_0$.

The main observation in this section is that all eigenfunctions of $H_S$ must be eigenfunctions of
different $J^\nu_0$, but with the same eigenvalue $k_0$. In contrast,
 all eigenfunctions of $H_\ell$ of the oscillator are eigenfunctions of
Casimir operator $Q$ for a fixed $q_\ell = \lambda_\ell(\lambda_\ell-1)$ and different eigenvalues
of $K_0(\alpha_\ell)$, namely 
$k_\nu=\nu+\lambda_\ell~, \nu=0,1,2,\cdots $.

\subsection{The traditional derivation of the bound state solutions} \label{trad}

For completness and for comparison, we review the traditional  derivation of
the bound state solutions \cite{morse,lal}. 
Making the change of variables
$$
  \xi = 2h(r) = 2k_0 e^{-(r-r_0)/a}~,
$$
in  Schr\"odinger's equation (\ref{sch}) and 
using $\Psi (r)=R(\xi)/r$, we obtain
\begin{equation} 
  R''(\xi) + \frac{1}{\xi} R'(\xi) + \left(-\frac 1 4 +
\frac{k_0}{\xi} + \frac{E}{{\cal E}\xi^2} \right)\,R (\xi)=0~. \label{R}
\end{equation} 
Again, substituting $ R(\xi)= e^{-\xi/2}\, \xi^s \,F(\xi)$ into (\ref{R}), where 
$s =\sqrt{-E/{\cal E}}$,
yields a differential equation for $F$ 
$$
F''(\xi) + (2s+1-\xi)\, F'(\xi)+ (k_0-1/2-s)\, F(\xi) =0~,
$$
whose solutions are the confluent hypergeometric functions $\,_1F_1(s+1/2-k_0,\, 2s+1;\, \xi)$.
These functions become polynomials and yield normalizable wavefunctions for 
$s=k_0-1/2-\nu \ge 0 $ where $\nu$ is nonnegative integer.
This condition yields the energy levels
$$
-E_\nu = {\cal E} s^2_\nu ={\cal E} (k_0-1/2-\nu)^2> 0~,\quad \nu=0,1,2,\cdots, \nu_{max} 
$$
and the corresponding wave functions
\begin{eqnarray} 
\Psi_\nu (r) &=& \frac 1 r ~R(2h(r)) \cr
&\propto& \frac 1 r e^{-h} h^{(k_0-1/2-\nu)}\,_1F_1(-\nu,\,2k_0-2\nu;\,2h)~.
\end{eqnarray} 
It is interesting to note that if $k_0=n+1/2$ we obtain $ E_n=0$, but this solution does not correspond to a bound state, since the solution is not normalizable for $s=0$ \cite{morse}. This result can be understood intuitively, because we are dealing 
with S-wave solutions, so that there is no potential barrier which can prevent the particle from escaping to infinity.
In contrast, an $E=0$ solution would probably be normalizble for
$\ell \ge 1$, since in the latter case the effective potential $U(r)= \ell(\ell+1)/(2Mr^2)+V_M(r)$
approaches $r \rightarrow \infty $ from above, and thus provides a potential barrier of infinite range.
This was illustrated by Daboul and Nieto \cite{dn}, who studied $E=0$ solutions for a class of potentials. 

\section{Effective Morse potentials generated by $J_\pm^\nu \equiv J_\pm(|E_\nu|/{\cal E})$}

Eqs.~(\ref{j012}) define a realization  $J_i(\beta)$ of the
${\mathfrak{so}}(2,1)\,$ algebra for every value of the parameter $\beta
\ge 0$. However, since we are interested in eigenfunctions of $J_0$ which
are normalizable, we retrict the values of $\beta$ to the discrete set
$\beta_\nu = -E_\nu/{\cal E}$\,, where the $E_\nu$ are the discrete
eigenvalues of $H_S$. We denote the corresponding generators by
\begin{equation}
J_i^\nu := J_i (\beta_\nu)=J_i (-E_\nu/{\cal E})\,, \quad \nu=0,1,2,\cdots
\label{knu}
\end{equation}
In (\ref{c1}) we found that
\begin{equation} 
J_0^\nu ~\Psi_{\lambda_\nu,\nu} (r) = k_0 ~\Psi_{\lambda_\nu,\nu} (r) 
\label{psi}
\end{equation}
Using the raising and lowering operators 
\begin{equation}
J_\pm := J_1 \pm i J_2\,,
\end{equation}
which obey
\begin{equation}
[J_0, J_\pm] = \pm  J_\pm\,,  \qquad
~  [J_+, J_-] = -2 J_0 \,,
\label{Kpm}
\end{equation}
the following functions can be defined, by  acting with 
$(J_\pm^\nu)^m$ onto $\Psi_{\lambda_\nu,\nu}$
\begin{equation}
\Psi_{\lambda_\nu,\nu \pm m} (r):= (J_\pm^\nu)^m ~\Psi_{\lambda_\nu,\nu} (r)\,.
\label{psim}
\end{equation}
If these states exist and are normalizable, then they must be eigenfunctions of $J_0$ with eigenvalues
$ k_0\pm m$, since
\begin{eqnarray}
J_0^\nu ~\Psi_{\lambda_\nu,\nu \pm m} (r) &=&
J_0^\nu ~(J_\pm^\nu)^m \Psi_{\lambda_\nu,\nu} (r) \cr
&=& (k_0 \pm m) (J_\pm^\nu)^m \Psi_{\lambda_\nu,\nu} (r) \cr 
&=& (k_0 \pm m)~\Psi_{\lambda_\nu,\nu \pm m} (r) 
\label{eigen}\end{eqnarray}
where we used the following general relations
$$
J_0^\nu ~(J_\pm^\nu)^m = (J_\pm^\nu)^m J_0^\nu +[ J_0^\nu,(J_\pm^\nu)^m]=
 (J_\pm^\nu)^m J_0^\nu \pm m (J_\pm^\nu)^m 
$$
Multiplying (\ref{eigen}) by the factor ${\cal E}\,2 h(r)$ and 
substituting the expression (\ref{hsj}) for the $J_0^\nu $ operator, 
yields the following differential equations
\begin{eqnarray}
0 &=& {\cal E} 2h(r) [J_0^\nu-(k_0\pm m)] ~\Psi_{\lambda_\nu,\nu \pm m} (r) \cr
&=& 
{\cal E} \left[- a^2\Delta_r - \frac{E_\nu}{\cal E} +  h^2(r) - 
(k_0 \pm m)2h(r)\right]~\Psi_{\lambda_\nu,\nu \pm m} (r)  \nonumber
\end{eqnarray}
The radial functions $\Psi_{\lambda_\nu,\nu+m} (r) $  can therefore be interpreted as 
eigenstates of the S-wave Schr\"odinger equation for the following potentials
\begin{eqnarray}
V_{\rm eff}(m,r) &=& \frac {\hbar^2}{2Ma^2} \left[ h^2(r) - (k_0+m)2h(r)\right]
 \cr
&=&  V_0 \left[e^{-2(r-r_0)/a}- 2\frac{k_m}{k_0} e^{-(r-r_0)/a} \right ]
\label{veff}\end{eqnarray}
where 
\begin{equation}
  k_m \equiv k_0 + m\,.
\label{km}\end{equation}
Each of these effective potentials has its minimum at $r_m$, where
\begin{equation}
   \frac{r_m}{a} = \frac{r_0}{a} - \ln\left(\frac{k_0 +m}{k_0}\right) =
\frac{r_0}{a} - \ln\left(\frac{k_m}{k_0}\right)\,, 
\label{rm} \end{equation}
so that 
\begin{equation}
e^{-(r-r_0)/a} = \frac{k_m}{k_0} ~e^{-(r-r_m)/a} ~.
\label{er} 
\end{equation}
By substituting this expression into (\ref{veff}), we obtain
\begin{eqnarray}
V_M(k_m,r_m; r) &=& V_m \left[e^{-2(r-r_m)/a}- 2 e^{-(r-r_m)/a} \right ]  \cr 
&=&  {\cal E}\,k_m^2 \left[e^{-2(r-r_m)/a}- 2 e^{-(r-r_m)/a} \right ]~. \label{Vm}
\end{eqnarray}
The effective potentials  (\ref{Vm}) look 
exactly as the original Morse potential (\ref{V}),
except that the parameters $(k_0,r_0)$ get changed into $(k_m,r_m)$.  
As $m$ increases the effective potentials (\ref{Vm}) will have shorter range and their minima
\begin{equation}
      V_{\rm eff}(r_m) =     - V_0\, \left(\frac{k_0+m}{k_0}\right)^2 =
-{\cal E}\,k_m^2  \label{depth}
\end{equation}
become deeper 
and deeper, decreasing almost quadratically with $m$.
The associated energy eigenvalues are
\begin{eqnarray}
 E_\nu^{(m)} &=& -{\cal E}\left(k_m-\frac{1}{2}-\nu\right)^2~, \quad \nu=0,1,\cdots, \lfloor k_m-1/2\rfloor_- \,, \cr 
&=&   -{\cal E}\left((k_0+m)-\frac{1}{2}-\nu\right)^2\, \cr
&=& E_{\nu+n}^{(m+n)}  \quad \mbox{for} \quad n\ge -\nu.  \label{enm}
\end{eqnarray}

\subsection{Connection to SUSY-QM} \label{sec:4a}

Using the relation (\ref{enm}) we obtain
immediately
\begin{eqnarray}
E_\nu &\equiv & E_\nu^{(0)}=-{\cal E}\left(k_0-\frac{1}{2}-\nu\right)^2=
-{\cal E}\left(k_{-\nu}-\frac{1}{2}\right)^2  \cr
&=& E_0^{(-\nu)} \, \quad \mbox{for}  \quad \nu=1,2,\cdots, \nu_{max} \,. 
\label{enmn} 
\end{eqnarray}
Hence the energies $E_\nu\,,~ \nu=1,2,\cdots, \nu_{max}$ of the excited states of the original Morse potential (\ref{V}) are equal to  {\em ground state} energies $E_0^{(-\nu)}$ of the effective potentials 
$ V_{\rm eff}(-\nu,r)~,~ \nu =1,2,\cdots,\nu_{max} $\,. 
This is one of the interesting results of the quantum-mechanical  
supersymmetry (SUSY-QM) formalism \cite{gms}. We derived it here without 
using the
latter formalism and without the need of finding out the relevant 
supersymmetric potential $W(x,a_i)$.

To understand the above result more thoroughly we give a second proof that the
$\Psi_{\lambda_\nu,\nu + m} (r)\equiv  (J_+^\nu)^m ~\Psi_{\lambda_\nu,\nu} (r)$ are eigenfunctions
of $H_S$ with the potential $V_{\rm eff}(m,r)$.

For this, we first note the important relation  (\ref{er}): it tells us that the $h(r)$ defined in (\ref{veff}) is invariant 
under the transformation $(k_0,r_0)$ to $(k_m,r_m)$, ~{\em i.e. } 
\begin{eqnarray}
h(r) &\equiv& h(k_0,r_0,r) = k_0 ~e^{-(r-r_0)/a} = k_m~e^{-(r-r_m)/a} \cr
&=& h(k_m,r_m,r)~.
\label{hrm} \end{eqnarray}
Hence, also the generators
\begin{equation}\label{jj}
J_i^\nu \equiv J_i(k_0, r_0,-E_\nu/{\cal E} ; r) = J_i(k_m, r_m,-E_\nu/{\cal E} ; r)\,,
\end{equation}
as defined in (\ref{j012}), do not depend on $m$, 
but differ for different $E_\nu/{\cal E}$.

Now we compare the following two equations : 
\begin{eqnarray}
&& J_0^\nu(k_0,r_0;r) ~\Psi_{k_0-\nu,\nu+m} (k_0,r_0;r) \cr
&&\qquad\qquad\quad = 
(k_0 + m)~\Psi_{k_0-\nu,\nu + m} (k_0,r_0;r) 
\label{eigen1} \\
&&J_0^\nu(k_m,r_m;r)~\Psi_{k_m-(\nu+m),\nu + m} (k_m,r_m;r) \cr 
&& \qquad\qquad \quad = k_{m}~\Psi_{k_m-(\nu+m),\nu + m} (k_m,r_m;r) µ~,
\label{eigenm}\end{eqnarray}
where the first follows from (\ref{eigen}) and the second follows from the condition 
(\ref{psi}) on the Morse eigenfunctions for the potential $V_{\rm eff}(m,r)$.
The two $\Psi$ are solutions of the same differential
operator $J_0^\nu$ with the same eigenvalue. Since the eigenvalues of $J_0^\nu$
are not degenerate, the two functions must be the same, except for a constant factor. 
Hence, $\Psi_{\lambda_\nu,\nu + m} (r)$ 
is the $(\nu + m)$-th excited state of the Morse potential $V_{\rm eff}(m,r)$.

\begin{figure}[htb]
\epsfig{file=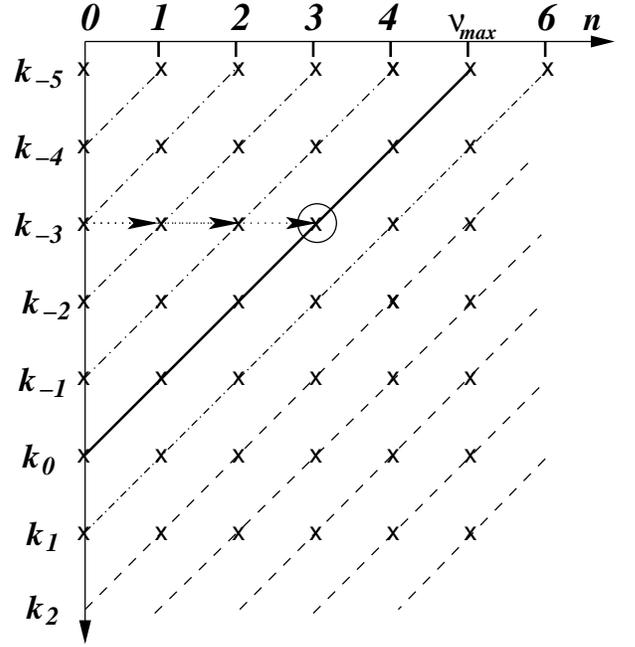,width=8.0cm,angle=0}
\caption
{The eigenstates for effective Morse potentials $V_{\!\!\rm eff}(m,r)~, m\ge -5$ are
displayed in the $(k_m, n)$ plane, where $k_m$ is related to the ground-state energies by 
$k_m=\sqrt{|E_0^{m}|/{\cal E}}+1/2$ and $n \equiv \nu\pm m$.
The eigenstates for a single  effective Morse potential $V_{\!\!\rm eff}(m,r)$
(\ref{veff}) are given along a diagonal line.  The number of bound states is given by 
$\lfloor k_m\rfloor_-+1 =\lfloor k_0\rfloor_- + m+1=\nu_{max}+m+1$.  
In particular, the original potential $V_M(r)$ in (\ref{V}), 
which corresponds to $m=0$, has $\nu_{max}+1$ bound states,
which lie along the solid diagonal line. Thus, for $m=-\lfloor k_0\rfloor_-$ 
there is only one bound state. In this diagram we chose $k_0=5.7$\, so that  $V_M(r)$ has
six bound states. In contrast to Fig.~\ref{fig:Fwy} all the states
with a fixed energy $E_\nu$ belong  to the same irreducible  representation of
${\mathfrak{so}}(2,1)\,$, and lie along  a horizonal line. 
The ground states for different $V_{\!\!\rm eff}(m,r)$ lie along the vertical line
$n=0$.   By moving along the horizonal lines we go from
eigenstates of one effective potential to another. This is illustrated by
the horizintal arrows. By applying $J_+^{\nu=-3}$ three times to the ground
state of $V_{\!\!\rm eff}(-3,r)~$, we obtain the third excited state of the original Morse
potential $V_M(r)$.
\label{diagram} 
}
\end{figure}

In Fig.~\ref{diagram}  we display the states $\Psi_{\lambda_\nu,n} (r)$,
defined in (\ref{psim}), in the $(\lambda_\nu, n)$ plane, where
$\lambda_\nu=k_0-\nu=\sqrt{|E_\nu|/{\cal E}}-1/2$ and $n \equiv \nu\pm m$.
All the eigenstates for a fixed effective potential $V_M(k_m,r_m; r)$ lie on
a single diagonal line. But, the states along a horizontal line have the
energy eigenvalue, but belong to different potentials $V_{\rm
eff}(m,r)$. The states $\Psi_{\lambda_\nu,\nu} (r)$ for the original Morse
potential (\ref{V}) can be obtained from the grounn states of $V_{\rm
eff}(-\nu,r)$ by applying $(J_+^\nu)^\nu$, as illustrted in the figure by
horizontal arrows.

\section{Summary and conclusions}
  The S-wave energy levels of the Morse potential have been known since 1929. Thus, the algebraic
derivation of these levels has not brought anything new, as far as applications are concerned.
What is interesting in the algebraic treatment is how the mathematical formalism works. In the
present paper we did clarify  when it works and how it works. 

We showed in particular that ${\mathfrak{so}}(2,1)\,$ is applied in a
completely different manner to the oscillator and the Morse potentials: for
the harmonic oscillator a single raising operator $K_+(\ell)$ yields all the
eigenstates for the given $\ell$. In contrast, for the Morse potential the
raising operators $J_+^\nu(\ell=0)$ map the wave functions of the original
Morse potential onto wavefunctions of other Morse potentials.  This
contrast is illustarted by figures 1 and 2. Thus, the message of the present
paper is that one should be more critical and check carefully how the SGA
are applied.

For example,  with our present insight we wanted to check how Wybourne
obtained the spectrum of the Morse potential in \cite[\S18.8]{wy}. It turned
out that Wybourne did not even give the ${\mathfrak{so}}(2,1)\,$ generators for
every potential, as was done, for example, in \cite{ch,cs}.  Instead, he used a slightly different version of the realization 
(\ref{kwy}) of ${\mathfrak{so}}(2,1)\,$ for the harmonic oscillator, and showed
that the algebra describes the energy spectrum of the following
differential equation~\cite[(Eq.(18.19)]{wy}
\begin{equation}
   \left(\frac{d^2}{dy^2} + \frac{a}{y^2} +by^2 + c\right)\,\Psi(y) = 0\,.
\end{equation}
Then he simply transformed the Schr\"odinger equation of different potentials to the above
differential equation. But this is essentially what Morse did, already in
1929, by transforming the S-wave Schr\"odinger equation to a diferential equation, which was
solved by Schr\"odinger. It is also what Landau and Lifschitz
\cite[\S23]{lal}  did for many potentials, by transforming 
their Schr\"odinger equations to the confluent hypergeometric equation. 

There is no doubt that group theory helps us understand many results in
physics, such as the degenercies of the eigenstates of the hydrogen atom. It
has many useful applications in elementary particles such as  flavor
and color SU(3). However, it seems to us that many of the papers on the SGA can
even mislead non-experts, as we demonstrated in the present paper.  They
might believe, for example, that one could obtain finite number of states by
just using the raising operators of a single realization of 
the ${\mathfrak{so}}(2,1)\,$  
algebra, as is the case for the oscillator.

Apparently, some experts have noticed that. In a well-written and easy to
read article \cite{gms}, the authors give a review of the super symmetry (SUSY-QM)
formalism and of shape invaraint potentials and mention that the
Morse potential is of the invariant type. This means that one can obtain the
excited states of the Morse potential from its ground state, not by applying
powers of a raising operator, as one naively expects, but by applying the
supersymmetric raising and lowering operators $A^\dagger(a_i)$ and $A(a_i)$ operators,
which have the same structure, but which depend on different parameters
$a_i$. We write this statement, using their notation, as follows
$$ 
\psi^{(-)}_{n+1}(x,a_0)  \sim  A^\dagger(a_1)A^\dagger(a_2)\cdots A^\dagger(a_n)\,\psi^{(-)}_{0}(x,a_n) 
$$
We showed in Sec.~\ref{sec:4a} that for the Morse  case the following equivalent statement holds
$$ 
\Psi_{k_0-\nu,\nu}(r,k_0)  \sim  (J_+^\nu)^\nu\,\Psi_{k_{-\nu},0} (r,k_{-\nu}) ~,
$$
where all the raising operators are equal, as we showed in (\ref{jj}).
Thus, we gave an explanation of why the excited states 
$\Psi_{k_0-\nu,\nu}(r,k_0)$ of
the original Morse potential (\ref{V}) are related to the ground states $\Psi_{k_{-\nu},0}(r,k_{-\nu})$
of related 
effective potentials $V$, without using the SUSY formalism.
We illustrate this action in Fig.~\ref{diagram} by applying $J_+^{\nu=3}$
three times on the ground state, $\Psi_{k_{-3},0}(r,k_{-3})$,  of $V(-3)$
and obtain the third excited state of $V(0)$.
A more recent and detailed review of the SUSY formalism can be found in \cite{kibler}.

It is interesting to note that by using a 'direct approach', it was possible
to obtain many results \cite{gomez}, among them a construction of new
quasi-exactly solvable deformation of the Morse potential, which the authors
have not been able to obtain by Lie-algebraic methods.

The Morse potential is
one of the simplest examples of the general class of Natanzon's potentials
\cite{natanzon} which have been studied by using SGA algebraic
methods~\cite{cs}.  Therefore  a deeper understanding of the SGA method in the Morse's case 
should help us understand the more general cases as well.

\begin{acknowledgments}

One of us (P.C.) acknowledges partial financial support from 
{\em Fondecyt} grant 1030993 and {\em Fondap} grant 11980002.
J. Daboul would like to thank Prof. Victor Fuenzalida, Chairman of 
the Physics Department, for his
hospitality and Dr. Juan Paulo  Wiff for his great help with the computer.
\end{acknowledgments}

\end{document}